\documentclass[prd, aps, superscriptaddress, preprintnumbers, twocolumn, floatfix, nofootinbib]{revtex4}

\usepackage{amsfonts}
\usepackage{amsmath}
\usepackage{amssymb}
\usepackage{bm}
\usepackage{dcolumn}
\usepackage{graphicx}   
\usepackage[latin1]{inputenc}
\usepackage{latexsym}
\usepackage{rotating}
\usepackage{hyperref}
\usepackage{graphicx}
\usepackage{color}
\usepackage{soul}

\usepackage{amsfonts}
\usepackage{amsmath}
\usepackage{amssymb}
\usepackage{bm}
\usepackage{dcolumn}
\usepackage{graphicx}
\usepackage[latin1]{inputenc}
\usepackage{latexsym}
\usepackage{rotating}
\usepackage{hyperref}

\newcommand\be{\begin{equation}}
\newcommand\bea{\begin{eqnarray}}
\newcommand\ee{\end{equation}}
\newcommand\eea{\end{eqnarray}}

\def\ba{\begin{array}}
\def\ea{\end{array}}

\renewcommand{\a}{\alpha}

\newcommand{\pa}{\partial}

\begin{document}

\title {Back-Reaction of Super-Hubble Cosmological Perturbations Beyond Perturbation Theory}

\author{Robert Brandenberger}
\affiliation{Physics Department, McGill University, Montreal, QC, H3A 2T8, Canada;\\
email: rhb@physics.mcgill.ca}

\author{Leila L. Graef}
\affiliation{Departamento de Fisica Teorica, Universidade do Estado do Rio de Janeiro, 20550-013 Rio de Janeiro, RJ, Brazil;\\
email: leilagraef@gmail.com}

\author{Giovanni Marozzi}
\affiliation{Dipartimento di Fisica, Universit\`a di Pisa, 
and INFN, Sezione di Pisa, Largo B. Pontecorvo 3, I-56127 Pisa, Italy,\\
and Centro Brasileiro de Pesquisas Fisicas, Rua Dr. Xavier Sigaud 150, Urca, CEP 22290-180, Rio de Janeiro, RJ, Brazil;\\
email: giovanni.marozzi@unipi.it}

\author{Gian Paolo Vacca}
\affiliation{INFN - Sezione di Bologna, via Irnerio 46, 40126 Bologna, Italy;\\
email: Gianpaolo.Vacca@bo.infn.it}

\date{\today}

%%%%%%%%%%%%%%%%%%%%%%%%%%%%%%%%%%%%%%%%%%%%%%%%%%%%%%%%%%%%%%%%%%%%%%%%%%%%%%%%%%%%%%%%%%%%%%
\begin{abstract}

We discuss the effect of super-Hubble cosmological fluctuations on the locally
measured Hubble expansion rate. We consider a large bare cosmological
constant in the early universe in the presence of scalar field matter
(the dominant matter component), 
which would lead to a scale-invariant primordial spectrum of cosmological
fluctuations. Using the leading order gradient expansion
we show that the expansion rate measured by a (secondary)
clock field which is not comoving
with the dominant matter component obtains a negative contribution from
infrared fluctuations, a contribution whose absolute value increases in time.
This is the same effect which a decreasing cosmological constant would produce.
This supports the conclusion that infrared fluctuations lead to a dynamical relaxation
of the cosmological constant. Our analysis does not make use of any
perturbative expansion in the amplitude of the inhomogeneities.

\end{abstract}
%%%%%%%%%%%%%%%%%%%%%%%%%%%%%%%%%%%%%%%%%%%%%%%%%%%%%%%%%%%%%%%%%%%%%%%%%%%%%%%%%%%%%%%%%%%%%%

\maketitle

%%%%%%%%%%%%%%%%%%%%%%%%%%%%%%%%%%%%%%%%%%%%%%%%%%%%%%%%%%%%%%%%%%%%%

%%%%%%%%%%%%%%%%%%%%%%%%%
\section{Introduction} 
%%%%%%%%%%%%%%%%%%%%%%%%%

The cosmological constant problem (see \cite{CCrevs} for reviews) is a key challenge for
fundamental physics. Basic arguments imply that the vacuum energy of matter fields
should act as a cosmological constant $\Lambda$ and cause the universe to accelerate. Assuming
that an ultraviolet cutoff scale close to the Planck scale is used, the resulting value for
$\Lambda$ is about 120 orders of magnitude larger than the maximal one allowed by
observations. The discovery of the acceleration of the Universe \cite{SN} has added a new perspective to
this problem. If the observed acceleration is due to a small cosmological
constant, then we do not only have to explain why the cosmological constant is not of
the Planck scale, but also why it happens to be rearing its head at the present time in
the cosmological history of the universe. This is the coincidence problem (see e.g.
\cite{DErevs} for reviews of the dark energy problem).

It has been conjectured for some time that de Sitter space is unstable because of infrared
effects \cite{Polyakov, Mottola} (see, however, \cite{Ian} for arguments
supporting the stability of de Sitter space), and that hence the bare cosmological constant in
the Lagrangian would be invisible today \footnote{See also \cite{Swamp, Dvali, Nima}
for a discussion on the difficulty of obtaining de Sitter space in string theory.}. 
Specifically, it was suggested by Tsamis and Woodard \cite{TW1}
that the back-reaction of super-Hubble scale gravitational waves could give a negative
contribution to the effective cosmological constant and cause the latter to relax. 
The problem was studied in perturbation theory, and it was found that one needs to
go to two-loop order (fourth order in the amplitude of the gravitational waves) 
in order to obtain a non-vanishing effect. The study of the
back-reaction effect of long wavelength cosmological perturbations was initiated
in \cite{ABM} and it was found that at one loop order (second order in the amplitude
of the perturbations) super-Hubble cosmological
perturbations lead to a negative contribution to the cosmological constant 
(see, also, \cite{Finelli:2001bn, Finelli:2003bp, Marozzi:2006ky}). Based
on this analysis, it was then conjectured in \cite{RHBbrRev} that this back-reaction
could lead to a late time scaling solution for which the contribution of the cosmological
constant tracks the contribution of matter to the total energy density. 

The back-reaction effect originates from the fact
that the Einstein equations are highly nonlinear. Hence, if we consider only linear
perturbations about a homogeneous and isotropic cosmological background
metric, then the Einstein equations are not satisfied at quadratic order from a naive 
point of view.  Then, specifically,
each Fourier mode of the linear fluctuations yields a contribution to the
background metric at quadratic order, and hence affects quantities such as
the Hubble expansion rate \footnote{In this approach second order perturbations, 
which would have a non-zero average, 
are incorporated in the left hand side of the Einstein equations producing, 
for example, an effective Hubble expansion. In this way, one takes into account the 
back-reaction, which can be evaluated looking at the aforementioned effect of linear 
perturbations at quadratic order.}.
Nonlinear effects also cause a correction to the
fluctuations themselves, but initial studies \cite{Patrick} have shown that these
effects are less important than the effects on the background.

The setup of the back-reaction analysis of \cite{ABM} was the following. A large
bare cosmological constant $\Lambda$ will lead to a phase of inflationary
expansion of space. Quantum fluctuations in this quasi de Sitter phase are 
continuously stretched beyond the Hubble radius, freeze out, are squeezed
and will generate an increasing phase space of super-Hubble modes. 
In terms of comoving momenta, the phase space of these infrared modes runs from
some value $k_i$ which corresponds to the Hubble radius at the initial time $t_i$
(and represents a physical infrared cutoff) to the Hubble scale $H e^{H(t - t_i)}$.
The fact that new modes are continuously injected into the phase space of
infrared modes is crucial for the back-reaction to be effective. Once the
back-reaction effect of the long wavelength modes has built up sufficiently,
it can cancel out the effects of the bare cosmological constant and terminate
the phase of accelerated expansion.

Questions about
the analysis of \cite{ABM} were raised in \cite{Unruh} where the challenge was posed
to show whether the effects predicted in \cite{ABM} are in fact locally measurable.
In fact, it was then shown in \cite{Ghazal1} (see also \cite{AW}) that in the case of
pure adiabatic fluctuations the effects computed in \cite{ABM} can be undone by
a second order time reparametrization. On the other hand, if there is a clock
field present in addition to the matter which dominated the energy density, then,
as shown in \cite{Ghazal2}, the back-reaction of cosmological fluctuations can be
shown to influence the locally measured expansion rate in the sense that the
locally measured expansion rate is smaller at a fixed value of the clock field
if there are super-Hubble fluctuations present than if there are none (assuming
that the clock field does not track the dominant component of matter, i.e. that
entropy fluctuations are present). Considering two components of matter,
a first dominant fluid (which sets up the cosmological fluctuations) and a second
sub-dominant clock field is very natural in the context of late time cosmology
where we measure time in terms of the temperature of the sub-dominant
radiation fluid. The analysis of \cite{Ghazal2} was extended in \cite{Marozzi:2012tp}
following a new gauge-invariant approach introduced in \cite{Finelli:2011cw}
(and based on \cite{Gasperini:2009wp, Gasperini:2009mu, Marozzi:2010qz}). 
This approach was first applied to analyze at second order in the perturbative 
expansion single scalar field models, for which only expansion rates defined by 
isotropic observers experience a non trivial negative quantum back-reaction~\cite{Marozzi:2011zb}.
Moving to the above mentioned two field models, in  \cite{Marozzi:2012tp}, it was in fact 
confirmed that , at second order in perturbation theory, the back-reaction of long 
wavelength cosmological
perturbations leads to a decrease in the locally measured expansion
rate (see also \cite{other} for other studies demonstrating that back-reaction
effects are for real). Furthermore, in \cite{Lam} it was shown physically how super-Hubble fluctuation modes
can modify the parameters of a local Friedmann cosmology.

All analyses of the back-reaction of cosmological fluctuations performed so far
are, however, analyses in leading order perturbation theory. Then, in almost the totality 
of the case back-reaction
effects of long wavelength fluctuations become important only
when perturbation theory breaks down (see, for example, \cite{Marozzi:2011zb})
\footnote{See \cite{Marozzi:2014xma} for a case in which back-reaction of tensor modes can be important 
within the perturbative regime, and \cite{Finelli:2008zg, Finelli:2010sh} for a first study of 
the regime of validity of perturbation theory through gauge invariant variables. Tensor modes back-reaction in a de Sitter background was 
also considered in \cite{Finelli:2004bm}.}.
Thus, while the fact that
the leading order perturbative back-reaction effect leads to a negative contribution
to the cosmological constant supports the possibility of an instability of (quasi) de
Sitter space-time, it cannot give a definite answer since the effect may be
undone by higher order effects. 

In this paper, we show that the back-reaction effect of super-Hubble cosmological
fluctuations on the local expansion rate persists beyond perturbation theory
and that, given fluctuations in the clock field relative to those of the
dominant matter field, the locally measured Hubble expansion rate obtains
a negative contribution, a contribution whose amplitude grows in time.
This supports the claim that (quasi) de Sitter space-time is unstable, and that it will
lead to a dynamical relaxation of the cosmological 
constant\footnote{Our qualitative result has been obtained 
quite simply, but with a little price, 
since there is a residual gauge dependence of the result because of our choice of 
separating average null non-homogeneous fluctuations. Given that our previous 
results in a fully gauge invariant framework are compatible with this assumption at 
second order in perturbation theory, we consider our analysis as a qualitative prediction 
at the non perturbative level. We shall address the problem of a fully non perturbative 
gauge invariant analysis in a future investigation.}.

Although our analysis is non-perturbative in the amplitude of the cosmological
perturbations, it is only a leading order analysis in the gradient expansion.
Such an expansion should be reasonable for super-Hubble fluctuations, and
the formalism we are using here was in fact suggested in \cite{Afshordi}
and applied to show that there is no parametric resonance of super-Hubble
scale metric fluctuations during reheating (see e.g. \cite{RhReview} for a
review) in the case of pure adiabatic perturbations. The analysis of
\cite{Afshordi} also shows that in the case of pure adiabatic fluctuations
there can be no back-reaction of super-Hubble fluctuation modes when 
the adiabatic field is the clock of the problem. However one obtains a non 
zero back-reaction from adiabatic fluctuation for an isotropic clock \cite{Marozzi:2011zb}.
As said, applying
 the formalism of \cite{Afshordi}, we here demonstrate that in the
presence of fluctuations of the clock field relative to the constant energy
density hypersurfaces of the dominant matter there is a non-vanishing
back-reaction effect, and that this effect corresponds to a negative
contribution to the locally measured Hubble expansion rate, a contribution
whose absolute value increases in time. 

The article is organized as follow.
In Section II we derive an expression for the local Hubble expansion
rate in terms of the variables expressing the cosmological perturbations.
We are interested in comparing the average of the expansion rate
taken over a hypersurface of constant clock field between a manifold with
cosmological perturbations and one without, at the same value of the clock
field. In Section III we evaluate this average expansion rate in the leading order
spatial gradient expansion (which will be a good approximation to study
the effects of super-Hubble fluctuation modes) and show that the fluctuations
result in a negative contribution $\Delta H$ to the expansion rate $H$ which corresponds to
a decrease in the effective cosmological constant. Let us also underline that, since $|\Delta H|$ is an
increasing function of time, the back-reaction effect corresponds in fact
to an instability of (quasi) de Sitter space and not just to a renormalization of the
cosmological constant. Finally, we conclude in Section IV.

%%%%%%%%%%%%%%%%%%%%%%%%%
\section{Local Hubble Expansion Rate} 
%%%%%%%%%%%%%%%%%%%%%%%%%

In this section we will derive an expression for the local Hubble expansion rate
from the point of view of a clock field $\chi$, extending the results of~\cite{Marozzi:2012tp}. 
This is determined in terms of
the normal vector $n_{\mu}$ to the constant $\chi$ hypersurfaces:
\be \label{2.1}
n_\mu \, = \, \frac{\partial_\mu \chi} {\sqrt{\xi}} \, ,
\ee
with $\xi\equiv g^{\alpha\beta} \partial_\alpha \chi \partial_\beta \chi$.
This can then be used to determine the local expansion rate $\theta(x)$, where $x$
denote the spatial coordinates, by
\be \label{Theta}
\theta \, = \, \nabla_\mu n^\mu= n^\mu \partial_\mu \log{\sqrt{-g}} + \partial_\mu n^\mu
\ee 
Starting from this expression, and following \cite{Finelli:2011cw},
we want than to compute the average of $\theta/\sqrt{Z_{\chi}}$ over
the constant $\chi$ hypersurface,  which will be used to define our effective 
expansion rate including the contribution of the fluctuations. 
Such an average can be defined by \cite{Finelli:2011cw}
\be
\langle A \rangle_\chi \, = \, \frac{\int \sqrt{-\bar{\gamma}} \, \bar{A}} {\int \sqrt{-\bar{\gamma}} } \, ,
\label{LER}
\ee
 where the barred coordinate system $\bar{x}^{\mu}=
 ({\bar{t}}, {\vec{\bar{x}}})$ is the system where $\chi$ is homogeneous, and ${\bar{\gamma}}$ is the determinant of the induced metric on that
 hypersurface.

Note that we treat $\chi$ as a spectator field in the same way that the radiation field
in late time cosmology can be viewed as a spectator field in the matter-dominated
phase. 
The dominant matter component can be modelled as a different scalar field
$\varphi$ which sets up the cosmological fluctuations. In the same way that
in current cosmology the dominant matter component is inhomogeneous from the
point of view of the constant radiation temperature surfaces, we assumed that the
dominant matter is inhomogeneous from the point of view of the constant $\chi$
surfaces. It is the dominant matter field which determines the metric fluctuations.
To describe these fluctuations we will work in generalized longitudinal gauge (LG)
(see e.g. \cite{MFB} for an in-depth review of the theory of cosmological
perturbations and \cite{RHBfluctRev} for a brief overview) in terms of which the
metric is given by
\be \label{metric}
g_{\mu\nu}={\rm diag}(e^{2\phi(t)}, - e^{-2\psi(t)}, - e^{-2\psi(t)},- e^{-2\psi(t)})\,,
\ee
where $\phi$ and $\psi$ are functions of space and time. In linear perturbation
theory, $\psi = \phi$ in the absence of anisotropic stress. Beyond linear perturbation
theory, however, $\phi$ and $\psi$ must be treated as independent.
Note
that this is the unique gauge in which the metric has no off-diagonal components.
If we want the metric to locally look like a Friedmann metric for long wavelength
fluctuations, then this gauge is the preferred one (see also~\cite{Marozzi:2011zb}).

The coordinate transformation from longitudinal gauge to the constant $\chi$
gauge is \cite{Marozzi:2012tp}
\bea 
x^\mu=(t,\vec{x}) \, \rightarrow \, \bar{x}^\mu \, &=& \, (\bar{t}, \vec{\bar{x}})  \\
&=& \, (\chi(t,\vec{x}),\vec{x}) \, \equiv \, f^\mu(x^\nu) \, , \nonumber 
\eea
and the metric in these coordinates, expressed in function of LG variables, becomes 
\begin{widetext}
\be \label{gtilde}
\bar{g}_{\mu\nu}(x) \, = \, \frac{1}{(\frac{\partial \chi}{\partial t})^2}
 \left(
\ba{cc}
e^{2\phi} & -e^{2\phi} \pa_i \chi   \\[10pt]
 -e^{2\phi} \pa_j \chi  &e^{2\phi}  \partial_i \chi   \partial_j \chi -e^{-2\psi} \delta_{i j}(\frac{\partial \chi}{\partial t})^2
 \ea
\right) \,,
\ee
with its inverse being
\be \label{gtildeinv}
\bar{g}^{\mu\nu}(x) \, = \, 
  \left.\left(
\ba{cc}
(\frac{\partial \chi}{\partial t})^2 e^{-2\phi} -|\vec{\nabla} \chi|^2 e^{2\psi} & -e^{2\psi} \vec{\nabla} \chi   \\[10pt]
 -e^{2\psi} (\vec{\nabla} \chi)^t  & -e^{2\psi}  \mathbb{I}
 \ea
\right)\right |_{f^{-1}(x) }=   \left. \left(
\ba{cc}
e^{-2\phi} & -e^{2\psi} \vec{\nabla} \chi   \\[10pt]
 -e^{2\psi} (\vec{\nabla} \chi)^t  & -e^{2\psi}  \mathbb{I}
 \ea
\right)\right |_{f^{-1}(x) }\,,
\ee
\end{widetext}
where all the quantities on the right hand sides are evaluated at $(f^{-1})^\mu(x^\nu)$.
Above we have used the fact that 
\bea
\xi \, 
&=& \, e^{-2\phi} (\frac{\partial \chi}{\partial t})^2 - e^{2\psi}(\vec{\nabla}\chi)^{2}  \nonumber
\eea
under the gauge transformation becomes
\be
{\bar \xi}(x) \, = \, e^{-2 \phi(f^{-1}(x))} \, .
\ee
Hence, the induced metric on the constant $\chi$ surfaces becomes
\be
ds^2= e^{2\phi}\frac{ (d\vec{x} \cdot \vec{\nabla} \chi  )^2}{\dot \chi^2}-e^{-2\psi} d\vec{x}^2 \, .
\ee
From (\ref{gtilde}) it follows that the determinant of the metric takes the form
\be
\sqrt{-\bar{g}} \, = \, \left( \frac{1}{\frac{\partial \chi}{\partial t}} e^{\phi -3 \psi}\right)_{f^{-1}(x)} .
\label{dettg}
\ee

We will now compute the local measure of expansion directly in the barred coordinates, i.e.
using
\be
\bar\theta(x) \, = \, \bar \nabla_\mu \bar n^\mu \, = \, 
\bar n^\mu \partial_\mu \log{\sqrt{-\bar g}}+\partial_\mu \bar n^\mu\, ,
\ee
where $\bar n^{\mu}$ is the normal vector to the constant $\chi$ hypersurfaces in
the barred coordinates.  In the following, all quantities which define the barred variables 
are evaluated at $f^{-1}(x)$.
We then have
\be
\bar{n}_\mu(x) \, = \,  \left[ {\pa x^\a\over \pa f^\mu}\right]_{f^{-1}(x)} n_\alpha\left(f^{-1}(x)\right) \,
\ee
which becomes
\be
\bar{n}_\mu(x) \, = \, 
\frac{1}{\sqrt{\xi(f^{-1}(x))}} \left(
\ba{cc}
\dot \chi  & \vec{\nabla} \chi
\ea
\right) 
\cdot
 \left(
\ba{cc}
1/\dot \chi   & -\vec{\nabla} \chi / \dot \chi    \\[10pt]
\vec{0}^t &  \mathbb{I}
\ea
\right) 
\ee
and, finally, gives
\be
\bar{n}_\mu(x) \, = \, 
 e^\phi(1,\vec{0}),
\ee
corresponding indeed (see Eq.~\eqref{2.1}) to the case of a homogeneous 
$\bar \chi$ which labels the time coordinate. One also has 
\be
\bar n^\mu \, = \, \bar g^{\alpha \mu} \bar n_{\alpha} \, ,
\ee
where the inverse metric is given in Eq.~\eqref{gtildeinv} so that
\be
\bar n^{\mu}=(1, -e^{2\psi} (\vec{\nabla} \chi)^t)|_{f^{-1}(x)} \,.
\ee

We also have  from Eq.~\eqref{dettg}
\be
\log{\left(\sqrt{-\bar g}\right)}=\phi-3\psi -\log{\dot \chi} 
\ee
which implies
\be
\pa_\mu \left[ \log{\left(\sqrt{-\bar g}\right)}\right]_{f^{-1}(x)} \, = \,  
\left(\pa_\mu (\phi-3\psi) +\frac{\pa_\mu \dot \chi}{\dot \chi} \right).
\ee
In the leading order gradient expansion the spatial derivative terms in the
above are negligible. Making use of $\bar n^{0} = exp(- \phi)$ we then get
\bea
\bar\theta(x) \, &=& \, \bar{g}^{00} \left( \dot\phi-3\dot\psi-\frac{\ddot \chi}{\dot \chi}\right) \\
& & + \bar{g}^{0i} \pa_i \left(  \phi-3\psi-\log{\dot \chi} \right)+\partial_\mu \bar g^{0\mu} \,  ,\nonumber
\eea
which then yields
\be
\bar\theta(x) \, = \, - 3 e^{- \phi} {\dot{\psi}} \, .
\ee

In order to compute the spatial average of ${\theta}$ over the constant $\chi$
hypersurfaces we need the determinant of the induced metric $\bar \gamma_{ij}$ on
these surfaces (see (\ref{LER})). This metric is obtained from the spatial part of 
(\ref{gtilde})
\be
\bar \gamma_{ij} \, = \, \frac{e^{2\phi}}{\dot \chi^2}  (\vec{\nabla} \chi)^t   \vec{\nabla} \chi -e^{-2\psi} \mathbb{I}
\ee
so that the measure factor is given by
\be \label{ind-det}
\sqrt{-\bar\gamma} \, = \, e^{-3\psi}\sqrt{1-e^{2(\phi+\psi)}\frac{ | \vec{\nabla} \chi |^2}{\dot \chi^2} } \, .
\ee
In the leading order gradient expansion we can neglect the spatial gradient
terms, and hence the above expression reduces to
\be \label{ind-det-2}
\sqrt{-\bar\gamma} \, = \, e^{-3\psi} \, .
\ee

We want now to compute the effective expansion rate $H_{eff}$. 
As introduced before, this can be defined as 
\be
H_{eff} \, = \, \dot{\chi}^{(0)} \frac{1}{3} \langle \frac{\theta}{\sqrt{\xi}} \rangle \, .
\ee
Noting that
$\bar \xi=e^{-2\phi}$, we would have at a classical level
\be
H_{eff}= \frac{1}{\int dx \sqrt{\bar\gamma}(x) } 
\int dx \sqrt{\bar\gamma(x)} e^{\phi(f^{-1}(x))} \bar\theta(x) \, .
\ee
where, in the equation above, we have considered $\dot{\chi}=1$.

\section{Evaluation of the Local Hubble Expansion Rate}

As we have seen in the previous section, the effective Hubble parameter is given by
\begin{equation}\label{heff}
H_{eff} \, = \, 
\frac{1}{3} \frac {\left< -3 e^{-3\psi_T}  \dot{\psi_T} \right>}{\langle  e^{-3\psi_T}\rangle } \, ,
\end{equation}
where $\psi_T$ is what we called $\psi$ in the previous section. The reason for this
change in notation is that in the following we want to denote by $\psi$ the fluctuating
part of $\psi_T$.

We can separate the background contribution from the total $\psi_{T}$ by writing
\begin{equation}
\psi_{T} \, = \, -ln(a/a_{0}) + \psi \, .
\end{equation}
There are two ways to set up this separation. In the first, we take $a(t)$ to be
a solution to the Friedmann equations in the absence of fluctuations. In this
case, the spatial average of the fluctuation $\psi$ only vanishes at linear order
in perturbation theory, but not at higher order. This is the view which was taken
in \cite{Marozzi:2012tp}. Here, on the other hand, we consider all contributions to the metric
which are homogeneous in space to be part of $a(t)$, or, more generally, to be part of the 
observable that we want to study including the back-reaction of the metric perturbation 
which have a non-zero average. In this manuscript, we consider as our observable the 
one defined in (\ref{heff}).
Hence, the spatial average
of $\psi$ vanishes even beyond linear order in perturbation theory. On the other hand,
the scale factor $a(t)$ will not naively satisfy the Friedmann equations if fluctuations are
present. The clock field $\chi$ evolves according to the full metric, not according to
the metric which obeys the Friedmann equations in the absence of fluctuations. Hence,
we will consider the separation where the spatial average of $\psi$ vanishes 
\footnote{Note that this separation between background and perturbation has a subtle 
gauge dependence, so that using this definition one cannot make fully gauge-invariant 
statements. On the other hand, as we shall see, this choice has the advantage of giving 
easily access to some qualitative results at a non-perturbative level in the leading 
order of the gradient expansion and we believe that this approach can give a first 
indication of what the non-perturbative back-reaction is. 
A more rigorous, fully gauge invariant, calculation is left for future work.}

We then write Eq.~(\ref{heff}) as
\bea 
H_{eff} \, &=& \, \frac{1}{3} \frac{\left< +3  a^{3}e^{-3\psi}(H_{hom}-\dot{\psi})\right>} {\langle  a^{3}e^{-3\psi}\rangle } \nonumber \\
&=& \,  \frac{\left<   e^{-3\psi} H_{hom}\right>} {\langle  e^{-3\psi}\rangle } - \frac{\left<  e^{-3\psi} \dot{\psi}\right>} {\langle  e^{-3\psi}\rangle }, \label{heff2}
\eea
where $H_{hom}$ is the Hubble expansion rate in the absence of perturbations, from which it is evident that
in order to obtain the contribution to $H_{eff}$ from the cosmological fluctuations we 
must compute the quantity
\begin{equation} \label{generalH}
\Delta H_{eff} \, \equiv \,  - \frac{\left<    e^{-3\psi} \dot{\psi}\right>} {\langle  e^{-3\psi}\rangle }. 
\end{equation}

In the following we will evaluate the above expression for the cosmological background
of interest to us, namely an inflationary phase driven by a bare cosmological constant 
but in presence of a scalar field $\varphi$ supporting gauge invariant fluctuations. 
In the absence of fluctuations, the expansion is characterized by a constant Hubble
expansion rate $H$. Like in inflationary cosmology, we assume that the phase
of accelerated expansion begins at some time $t_i$, and we denote by $k_i$ the comoving
momentum whose wavelength is equal to the Hubble radius at the initial time.
Causal dynamics of the accelerated phase cannot determine anything about
fluctuations on larger length scales, and we will introduce a physical infrared cutoff
by setting any initial super-Hubble fluctuations to zero.

Let us begin by evaluating $\Delta H_{eff}$ to leading order in perturbation theory.
By expanding the term $e^{-3\psi}$ in the expression of $\Delta H_{eff}$ we obtain,
\be
\Delta H_{eff} \, \approx \,  - \frac{\left< (1-3\psi) \dot{\psi}\right>} {\langle (1-3\psi)\rangle } \, . 
\ee
The term linear in $\dot{\psi}$ vanishes when taking the 
average. Hence, we obtain
\begin{equation}\label{heffpert}
\Delta H_{eff} \, \approx \,  - \left< -3\psi \dot{\psi}\right>.
\end{equation}

We can Fourier expand the fluctuation $\psi(t,x)$
\begin{equation}
\psi(t,x) \, = \, \int d^{3}k \; \epsilon_{\bf k} \;  \psi_{k} \; e^{i {\bf k} x + \alpha(k)} \, 
\end{equation}
where $\psi_k$ represent the amplitudes of the modes (and hence are positive), $\alpha(\bf k)$
are phases, and the $\epsilon_{k}$ are independent Gaussian random variables, i.e.
\be
\left< \epsilon_{\bf k} \epsilon_{\bf k'} \right> \, = \, \delta({\bf k} - {\bf k}') \, .
\ee
Since the fluctuations produced during the inflationary phase have a roughly scale-invariant
spectrum \cite{ChibMukh}, we have (neglecting the tilt)
\begin{equation}
P(k) \, = \, |\psi_{k}|^{2} k^{3} \, = \, \rm{const} \, .
\end{equation}
Therefore
\begin{equation} \label{spectrum}
\psi_{k} \, \sim \,  k^{-3/2}.
\end{equation}

To obtain the back-reaction effect of super-Hubble modes, we must integrate over all
values of ${k}$ with $ k \equiv |{\bf k}|$ between the Hubble crossing scale and the
infrared cutoff $k_i$ described above. 
We obtain
\begin{equation}
3<\psi(t,x)\dot{\psi}(t,x)> \, = \, 3 \int d^{3}k \psi_{k} \dot{\psi}_{k} \, .
\end{equation}
To evaluate this term we use the results of the theory of cosmological
fluctuations which tells us (see e.g. \cite{MFB}) that $\psi_{k}$ is
constant on super-Hubble scales modulo a decaying mode. Thus, we 
can write $\psi_{k}$ as 
\begin{equation}
\psi_{k}\, = \, A_{0k} + \delta A_{k}(t) \, ,
\end{equation}
where $A_{0k}$ is the constant mode, and 
$\delta A_{k}(t)$ is the decaying mode which in an inflationary background
scales as
\be \label{decaying}
\delta A_{k} \, = \, C_7 A_{0k} e^{-H(t-t_{H}(k))} \,, 
\ee
where $t_{H}(k)$ is the time when the mode $k$ crosses the Hubble radius,
and $H$ is $H_{hom}$.
Since both modes have equal strength at Hubble radius crossing, the coefficient
$C_7$ is positive and its absolute value is of order one. Note that the 
fact that for a super-Hubble fluctuation the amplitude of the adiabatic
mode is a constant plus a decaying piece is valid also beyond the perturbative regime
(see e.g. \cite{LV}). Above, $A_{0k}$ is the contribution considered
previously (\ref{spectrum}), namely 
\be
A_{0k} \, \sim \,  k^{-3/2} \, , 
\ee
which was obtained since the spectrum of fluctuations produced during
the inflationary phase is (almost) scale invariant spectrum. From (\ref{decaying})
it immediately follows that
\begin{equation}
\dot{\delta A}= - C_7 A_{0k} H e^{-H(t-t_{H})}.
\end{equation}
Consequently,
\bea
 A_{0k}\dot{\delta A}_{k} \, &=& \, - C_7 A_{0k}^{2} H e^{-H(t-t_{H})} \\
 &=& \,  - C_8 \; k^{-3} H e^{-H(t-t_{H})} \, , \nonumber
\eea
where $C_8$ is another positive constant. Integrating this contribution
over all super-Hubble modes therefore yields
\begin{equation}
<\psi(t,x) \dot{\psi}(t,x)> \, = \,  - C_8 H e^{-Ht}\int^{He^{Ht}}_{k_{i}} dk k^{2} k^{-3}  e^{H t_{H}(k)} \, .
\end{equation}
Since $e^{Ht_{H}} = k/H$ we obtain
\bea \label{psidotpsi}
<\psi(t,x) \dot{\psi}(t,x)> \, &=& \, - C_8 e^{-Ht} H e^{Ht} \bigl( 1 - \frac{k_i}{H} e^{- Ht} \bigr) \nonumber \\
&=& \, - C_8 H \bigl( 1 - f(t) \bigr) \, ,
\eea
where $f(t)$ is a positive decreasing function of time (always smaller than 1).

Therefore, substituting the above equation into the 
expression for $\Delta H_{eff}$, given by (\ref{heffpert}) we obtain
\begin{equation}
\Delta H_{eff} \approx  - 3C_8 H(1 - f(t)) \, , 
\end{equation}
where $C_8$ is positive and $f(t)$ is a positive decreasing function of time.
We can see that in the perturbative regime the back-reaction effect of super-Hubble
cosmological fluctuations yields an increasingly negative contribution to the Hubble parameter.

Now let us analyze the effective Hubble parameter beyond perturbation
theory, but in leading order in the gradient expansion. We begin with the
previously derived general expression of $H_{eff}$ in Eq. (\ref{heff2}).
We consider the term
\begin{equation} \label{deltaH1}
\Delta H_{eff} \,  \equiv  \,  - \frac{\left< e^{-3\psi} \dot{\psi}\right>} {\langle  e^{-3\psi}\rangle } .
\end{equation}
At any time we can Fourier expand the fluctuation field $\psi(x, t)$ 
\be \label{gansatz}
\psi(x, t) \, = \, A(t) g(x, t) \, 
\ee
where $A(t)$ characterizes the amplitude of the fluctuation and $g(x, t)$ is a function of unit 
amplitude whose spatial average vanishes (since $\psi$ is a fluctuation whose spatial average
vanishes). Neglecting, for a moment, the fact that modes cross the Hubble radius,
it would then conclude from the conservation of adiabatic fluctuations on super-Hubble
scales (see e.g. \cite{LV}) that the amplitude $A(t)$ has a constant
component $A_0$ and a decaying piece $A_1(t)$, i.e.
\be \label{modes}
A(t) \, = \, A_0 + A_1(t) \, ,
\ee
both multiplying the same function $g(x, t)$. Recall that for each Fourier mode of the
fluctuation, the two modes have comparable amplitude when the mode exits
the Hubble radius, and the second mode afterwards decreases as $exp(-H(t - t_e))$,
where $t_e$ is the time when the mode exits the Hubble radius. 
If we neglect the fact that new modes cross the Hubble radius (i.e. neglecting the
increase of the phase space of super-Hubble modes) the function g(t,x) would be
independent of time. At first, we will work in an adiabatic approximation in which we
neglect the time-dependence of $g$. In evaluating (\ref{deltaH1}), we
make use of the fact that only the second mode in (\ref{modes})
depends on time, and that the overall amplitude of this mode remains
constant when we take into account that new modes are continuously
exiting the Hubble radius. Hence, $\dot{\psi} = - H A_1 g$, and
\be \label{deltaH1b}
\Delta H_{eff} \,  \equiv  \,  
H \frac{\left< e^{-3 A(t) g(x, t)} A_1 g(x, t) \right>} {\langle  e^{-3\psi}\rangle } \, .
\ee
The factor $e^{-3 A(t) g(x,t)}$ acts as a weighting function. It gives larger weight to
values of $x$ where $g(x, t)$ is negative. Hence, the expectation value in the
numerator of (\ref{deltaH1b}) is negative, and we conclude that
\be \label{result1}
\Delta H_{eff} \, < \, 0 \, .
\ee
Since the phase space of super-Hubble modes is increasing, the effective overall
value of $A$ will increase in time (because the phase space of infrared modes
is increasing). Hence, the absolute value of $\Delta H_{eff}^{(1)}$
will be increasing in time, i.e.
\be \label{result2}
\frac{d}{dt} \Delta H_{eff} \, < \, 0 \, .
\ee
Note that (\ref{result1}) and (\ref{result2}) are the same conclusions obtained in the 
perturbative analysis.

The inequalities in (\ref{result1}) and (\ref{result2}) are the main results of our analysis.
They demonstrate that, to leading order in the gradient expansion, the back-reaction
of super-Hubble cosmological fluctuations leads to a decrease in the expansion rate which an
observer described by a clock field $\chi$ (which has a negligible contribution to the
energy density) measures. The absolute magnitude of the back-reaction effect
increases in time as more modes become super-Hubble. The main new feature
of the present analysis (compared to previous work) is that our analysis does
not make use of a perturbative expansion in the amplitude of the cosmological
perturbations. We stress that a residual gauge dependence is present 
in this approach, which, we believe, is nevertheless catching at a qualitative level, 
with really minimal efforts, the right phenomenological behavior of the 
effect of the back-reaction.

We also remind the reader that we have used a {\it quasi-adiabatic} 
approximation in which we at any time $t$
consider the phase space of modes which are super-Hubble at time $t$, and treat
it as a time-independent phase space. In this approximation, we compute the
magnitude of the change in the Hubble expansion rate, finding
$\Delta H_{eff} < 0$. In a second step we
ask how the modes crossing the Hubble radius change the position space
amplitude of the fluctuation, and then reach the conclusion that the absolute
value of $\Delta H_{eff}$ increases in time. It would be nice to find an analysis
which avoids having to make this approximation.

%%%%%%%%%%%%%
\section{Discussion}
%%%%%%%%%%%%%

We have studied the effect of super-Hubble cosmological fluctuations on the locally
measured Hubble expansion rate. We worked in the leading order gradient
expansion, but nonperturbatively in the amplitude of the fluctuations 
as a novel step in our approach.
We consider a large bare cosmological constant which leads to 
accelerated expansion which in turn generates an (almost) scale-invariant spectrum of cosmological
fluctuations on super-Hubble scales. We have shown that the expansion rate measured by a 
clock field which is not comoving
with the dominant matter component obtains a negative contribution from
infrared fluctuations, a contribution whose absolute value increases in time.
This is the same effect which a decreasing cosmological constant would produce.
This supports the conclusion that infrared fluctuations lead to a dynamical relaxation
of the cosmological constant \cite{RHBbrRev}.

%%%%%%%%%%%%%%%%
\section*{Acknowledgement}
%%%%%%%%%%%%%%%%
\noindent

The research at McGill is supported in
part by funds from NSERC and from the Canada Research Chairs program.
L.L.G. is supported by a posdoc grant "Pos-Doutorado
Nota 10" from Fundacao Carlos Chagas Filho de Amparo
a Pesquisa do Estado do Rio de Janeiro (FAPERJ),
No. E - 26/202.511/2017.
GM wishes to thank  INFN, under the program TAsP (Theoretical Astroparticle Physics), and 
CNPq for financial support.

%%%%%%%%%%%%%%%%%%%

\end{document}